\documentstyle[epsf]{mn}
\input{epsf}
\def\lsim{\mathrel{\hbox{\rlap{\hbox{\lower4pt\hbox{$\sim$}}}\hbox{$<$}}}}
\def\gsim{\mathrel{\hbox{\rlap{\hbox{\lower4pt\hbox{$\sim$}}}\hbox{$>$}}}}

\title[Generalised Inflation Model]
{Generalised Inflation with a Gravitational Wave Background}

\author[Lukash et al.]  {V.N. Lukash$^{1}$,
       E.V. Mikheeva$^1$, V. M\"uller$^2$, and A.M. Malinovsky$^1$\\
       $^1$Astro Space Center of P.N.Lebedev Physical Institute \\
       Profsoyuznaya 84/32, 117810, Moscow, Russia\\
       $^2$ Astrophysikalisches Institute Potsdam,
        An der Sternwarte 16, Potsdam, D-14482 Germany}
\date{Accepted. Received; in original form}

\begin{document}
\maketitle

\begin{abstract}
We propose a $\Lambda$-inflation model which explains a significant part of
the COBE signal by primordial cosmic gravitational waves.  The primordial
density perturbations fulfill both the constraints of large-scale microwave
background and galaxy cluster normalization.  The model is tested against the
galaxy cluster power spectrum and the high-multipole angular CMB anisotropy.
\end{abstract}

\section{Introduction}

The observational reconstruction of the {\it cosmological density
perturbation} (CDP) spectrum is a key problem of modern cosmology.  It
provided a dramatic challenge after detecting the primordial {\it cosmic
microwave background} (CMB) anisotropy by DMR COBE (Smoot et al.  1992,
Bennett et al.  1992) as the signal found at $10^0$, $\Delta T/T = 1.06\times
10^{-5}$ (Bennett et al.  1996), appeared to be few times higher than the
expected value of $\Delta T/T$ in the most simple and best developed
cosmological model with standard {\it cold matter} (sCDM)\footnote{The 
matter density $\Omega_m =\Omega_b +\Omega_{cm} =1$, there is no cosmological
term , $h = 0.5$, the slope of the CDP spectrum $n_S = 1$, i.e.  of
Harrison-Zeldovich type, no contribution from cosmological gravitational
waves.}.
Indeed, the CDP spectrum of sCDM normalized by the biasing parameter $b^{-1}
\equiv \sigma_{8} = 0.6$ (White, Efstathiou, Frenk 1993, Eke, Coles, Frenk
1996, Viana, Liddle 1996) can reproduce only ~30\% of the COBE measured CMB
anisotropy.  If $\sigma_{8}$ is less than $0.6$, this contradiction gets even
stronger.

During recent years there were many proposals to improve sCDM\footnote{In the
simplest terms, to remove the discrepancy between the CDP spectrum amplitudes
at $8h^{-1} \rm{Mpc}$ as determined by galaxy clusters, and at large scales,
$\sim 1000h^{-1} \rm{Mpc}$, by the $\Delta T/T$ inhomogeneity.}  by adding
hot dark matter, a $\Lambda$-term, or considering non-flat primordial CDP
spectra.  Below, we present another, presumably more natural way to solve the
sCDM problem based on taking into account a possible contribution of {\it
cosmic gravitational waves} (CGWs) into the large-scale CMB anisotropy.
Further, we will also try to preserve the original near-scale-invariant CDP
spectrum in the CDM universe.  Thus, the problem is reduced to the
construction of a simple inflation producing a near {\it Harrison-Zel'dovich}
(HZ) spectrum of CDPs ($n_S\simeq 1$) and a large relative contribution of
CGWs into the $\Delta T/T$ at COBE scale (i.e.  the ratio of the tensor to
scalar mode contributions T/S $\sim$ 1).

A basic physical reason for the production of tensor and scalar perturbations
in the expanding Universe is the {\it parametric amplification effect}
(Grishchuk 1974, Lukash 1980):  the spontaneous creation of quantum physical
fields in a non-stationary gravitational background.  From the theoretical
point of view a cosmology with negligible contribution of CGWs may be
considered as a phenomenological model only, because any inflationary model
predicts a non-zero CGW amplitude.  Generally, the tensor mode is not
discriminated in inflation (e.g.  Starobinsky 1979, Rubakov, Sazhin,
Veryaskin 1982).  Different models with nearly scale-invariant spectra of
CDPs predict different abundance of CGWs (Lucchin, Matarrese 1985, Linde
1994, Garsia-Bellido, Linde, Wands 1996, Lukash, Mikheeva 1996).  However, in
chaotic and power-law inflations it is usually small, T/S ${}^{<}_{\sim}$ 1,
which corresponds to the spectrum slope $n\ge 0.8$.  An always `red' CDP
spectrum ($n_S < 1$) originating in the power-law inflation model helps to 
satisfy the `double-normalization' (i.e.  to reconcile the $\sigma_8$ {\it vs.}  
$\Delta T/T$ problem).

So, the first candidate of a successful model fitting both COBE and galaxy
cluster normalizations could be a power-law spectrum of CDPs with a slope $n_S
\simeq 0.85$ predicted by power-law inflation (e.g. Lucchin, Matarrese 1985, 
there an exponential potential of the inflation field is used). One has a 
simple estimate for the fraction of CGWs,
\begin{equation}
\frac{\rm T}{\rm S} \simeq -6n_{T},
\end{equation}
where $n_T$ is the slope of the CGW spectrum related trivially to $n_S$ in
the case of power-law inflation:  $n_T = n_S - 1$.

Notice that large T/S and hence the desired double-normalization can be
reached in this model only at the expence of a rejection from the HZ
spectrum:  the red power-law CDP spectrum helps on large scale, however it is
getting undesirable on Mpc scale producing a too late galaxy formation epoch
(Gardner et al.  1997, and references cited therein).

Therefore, it is interesting to consider other models with high abundance of
CGWs but at the same time with $n_S \simeq 1$ unchanged (with a preference of
HZ or a slightly blue CDP spectrum on short scales to provide early formation
of high-redshift quasars and early galaxy formation).

A simple model of such kind is $\Lambda$-inflation, an inflationary model
with an effective metastable $\Lambda$-term (Lukash, Mikheeva 1996, 1997,
2000, Mikheeva 1997).  This model produces both curvature and primordial
gravitational wave perturbations which have a non-power-law spectrum, with a
shallow minimum in the CDP spectrum, located at a scale $k_{cr}$ (where the
$\Lambda$-term and the scalar field have equal energy densities).  Around this
scale, the CDP spectrum is exactly of the scale invariant HZ form, and the
ratio T/S is close to its maximum; it is of the order unity depending on
the model parameters.

The estimate (1) remains true for $\Lambda$-inflation and, probably, keeps its
universality for any type of inflationary dynamics (however, the relationship 
between $n_T$ and $n_S$ can vary).  The cost to be paid for the possibility of
having a high ratio T/S on scales where $n_S\simeq 1$ is the non-power-law 
CDP global spectrum:  it is `red' at $k<k_{cr}$ and `blue' at $k>k_{cr}$.  
This smooth transition in the spectrum slope from red to blue makes T/S 
obviously dependent on how $k_{COBE}$ is related to $k_{cr}$.  One may benefit
from the HZ local slope or a blue spectrum enhancement on Mpc scales (to 
initiate the early structure formation) by simply adjusting $k_{cr}$ to a 
galaxy cluster scale.

\section{$\Lambda$-inflation with self-interaction}

We summarize briefly the main properties of $\Lambda$-inflation and 
determine the basic model for our investigation.  Let us consider 
a general potential of $\Lambda$-inflation:
\begin{equation}
V(\varphi) = V_0 + \sum_{\kappa=2}^{\kappa_{max}} 
\frac{\lambda_{\kappa}}{\kappa}\varphi^{\kappa},
\end{equation}
where $\varphi$ is the inflaton scalar field, $V_0 > 0$ and 
$\lambda_{\kappa}$ are constants, and  $\kappa=2, 3, 4,..$.

In the case of a massive inflaton ($\kappa=\kappa_{max} = 2$, $\lambda_2
\equiv m^2 > 0$, this model is called $\Lambda m$-inflation) T/S can be
larger than unity only when the CDP spectrum slope in the `blue' asymptote is
very steep, $n_S^{blue} > 1.8$.  To avoid such a strong spectral bend on
short scales ($k > k_{cr}$), we choose here another simple version of
$\Lambda$-inflation -- the case with self-interaction:  $\kappa=\kappa_{max}
= 4$, $\lambda_4 \equiv \lambda > 0$; this model is called
$\Lambda\lambda$-inflation.

The scalar field $\varphi$ drives an inflationary evolution if $\gamma 
\equiv - \dot H/H^2 < 1$, where\footnote{We assume Planckian units $8\pi 
G = c = \hbar = 1$, a dot denotes the time derivative ($\equiv d/ad\eta$), 
and $a$ and $H \equiv \dot a/a$ are scale and Hubble factors, respectively.}  
$H =\sqrt{V/(3-\gamma)} \simeq \sqrt{V/3}$.  This condition holds true for
all values of $\varphi$ (at worst, for $\varphi \sim \varphi_{cr}$, where
$V_0 = \lambda \varphi^4_{cr}/4$ and $\gamma$ reaches its maximum) if
\begin{equation} c \equiv \frac 14 \varphi_{cr}^2 =\frac 12
\sqrt{\frac{V_0}{\lambda}} > 1, \end{equation} 
which we imply hereafter.

The last inequality simultaneously ensures the validity of the slow-roll
approximation ($\vert \ddot \varphi/ \dot \varphi \vert \ll 3H$) as
\begin{equation}
\gamma \simeq \frac 2c \left(\frac{y^3}{1 + y^4}\right )^2,\;\;\;\;\;\;
\frac{\ddot\varphi}{3H\dot\varphi}\simeq -\frac{y^2 (3+y^4)}{3c(1+y^4)^2},
\end{equation}
where $y\equiv\varphi/\varphi_{cr}=\varphi/\left(2\sqrt c\right)$.  
Then we get the value of the inflaton field at horizon-crossing time 
($k=aH\simeq -1/\eta$),
\[
cy^2 = \sqrt{c^2 +x^2}-x,
\]
and the gravitational perturbation spectra $q_k$ and $h_k$ generated in 
S and T modes, respectively (see the Appendix):
\begin{equation} 
q_k = \frac H{2\pi\sqrt{2\gamma}} = \frac{\sqrt{2\lambda/3}}{\pi}
\left(c^2+x^2\right)^{3/4}, \end{equation} 
\begin{equation} 
h_k = \frac{H}{\pi\sqrt{2}} = \frac{2c\sqrt{\lambda/3}}{\pi} 
\left ( 1 + \frac x{\sqrt{c^2 + x^2}} \right)^{-1/2}, 
\end{equation} 
where 
\begin{eqnarray}
x &=&\ln\left[ \frac{k}{k_{cr}y} \left(\frac{2}{1+y^4}\right)^{1/6} 
\right] \nonumber \\
&=&\ln\left[\frac{k}{k_{cr}}\left(1+\left(\frac xc\right)^2\right)^{1/4} 
\left (1+\frac x{\sqrt{c^2+x^2}}\right)^{2/3}\right] \nonumber \\
&\simeq&\ln(k/k_{cr}).
\nonumber 
\end{eqnarray}
Fig.1 shows the power spectra for $c=5,9,11$. 

\begin{figure}
{\epsfxsize=9cm \epsfysize=8truecm
\epsfbox[25 10 226 255]{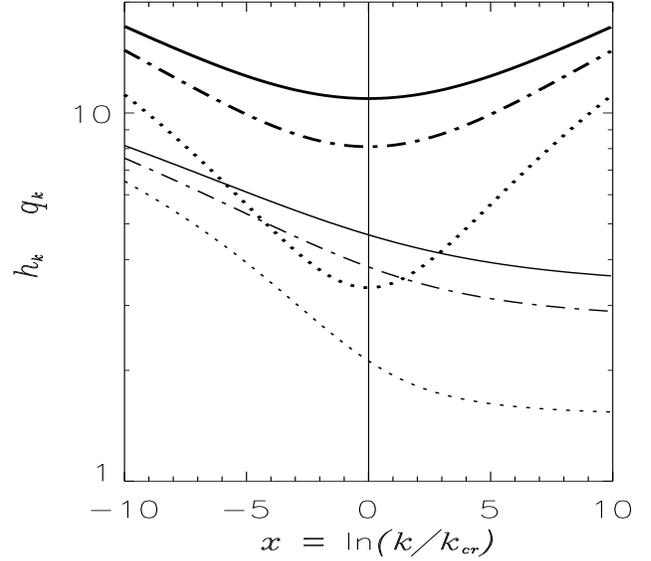}}
\vspace{0.2cm}
\caption{Spectra of curvature perturbations $q_k$ (thick lines) and 
gravitational waves $h_k$ (thin lines) around the typical scale $k_{cr}$ 
for $c=5,9,11$ as dotted, dash-dotted, and solid lines, respectively (arbitrary
normalization).}
\end{figure}

The dimensionless power spectrum of CDPs is directly related to the
fundamental $q_k$ spectrum
\begin{equation}
\langle~\delta^2~\rangle=\int\limits^\infty_0\Delta_k^2\frac{dk}k,\;\;\;\;
\Delta_k = 3.6\times 10^6 \left(\frac{k}{h} \right)^2q_k T(k),
\end{equation}
where the wave number $k$ is measured in $h/\rm{Mpc}$, and $T(k)$ is 
the transfer function.  

We can find also the local slopes of the fundamental power spectra 
(5), (6):
\begin{equation}
n_S -1 \equiv 2\frac{ d \ln q_k}{d \ln k} = \frac{3x}{c^2 +x^2}
\in \left[ -\frac{3}{2c}, \frac{3}{2c} \right],
\end{equation}
\begin{eqnarray}
n_T \equiv 2 \frac{d\ln h_k}{d\ln k} = -2 \gamma = -
\frac{1}{\sqrt{c^2 + x^2}} \left( 1- \frac{x}{\sqrt{c^2 + x^2}}
\right) \\
\in \left[-\frac{3\sqrt 3}{4c},0 \right]. \nonumber
\end{eqnarray}
Obviously, the maximal deviations from the HZ spectrum take place in 
S-mode at $x=\pm c$, and in T-mode at $x = -c/\sqrt 3$.  At the 
latter point the ratio of spectra reaches its maximum,
\begin{equation}
\left(\frac{h_k}{q_k}\right)^2 = 4\gamma = 2\vert n_T\vert\;
\le\;\frac{3\sqrt 3}{2c}.
\end{equation}

\section{CDM cosmology from $\Lambda$-inflation}

Let us consider the CDP spectrum $\Delta_k$ (7) with CDM transfer function,
normalized both at the large-scale $\Delta T/T\vert_{10^0}$ (including the
contribution from $h_k$ (6)) and the galaxy cluster abundance at $z = 0$
($\sigma_8 $), to find the family of the most realistic S-spectra $q_k$
produced in $\Lambda\lambda$-inflation.

In total, we have three parameters entering the $q_k$ spectrum:  $\lambda$
(the overall amplitude), $c$ (the measure of T/S) and $k_{cr}$ (the scale
where the CDP spectrum is locally HZ, $n_S=1$\footnote{$n_S=1$ is also in the
asymptotics $\vert x \vert \gg c$.}).  Constraining them by two observational
tests, we are actually left with only one free parameter (say, $k_{cr}$)
which may be restricted elsewhere by other observations (e.g.  cluster power
spectra, acoustic peaks, bulk velocities, etc.).

To demonstrate explicitly how the three parameters are mutually related, we
first employ simple analytical estimates for the $\sigma_8$ and $\Delta T/T$
tests to derive the key equation relating $c$ and $k_{cr}$, and then solve
this equation numerically to obtain the range of interesting physical
parameters.

Instead of taking the $\sigma$-integral numerically ($R=8h^{-1}\rm{Mpc}$),
\begin{equation}
\sigma^2_R = \int \limits^\infty_0 \Delta^2_k W^2 (kR)\frac{dk}k,\;\;\;\;
W(z)=\frac 3{z^3} (\sin z -z\cos z),
\end{equation}
we may estimate the wavelength $k_1$ at which $\Delta_{k_1} = 1.6 \sigma_8$
(to be equal unity for $\sigma_8 = 0.6$)\footnote{The integral is roughly
estimated as $\sigma^2_8\simeq \Delta^2_{k_1}/\alpha$ assuming that the
integrand grows sharp with $k$ and $\Delta^2_k\sim k^{\alpha}$ near $k\sim
k_1$.  For sCDM, $\alpha\simeq 2.5$ at $k_1\simeq 0.3h/ \rm{Mpc}$.}. This
will fix the spectrum amplitude on the cluster scale (see eq.(7)):
\begin{equation}
q_{k_1} \simeq 4.5\times 10^{-7}\frac{h^2 \sigma_8}{k_1^2 T(k_1)}\;\;.
\end{equation}

On the other hand, the spectrum amplitude on large scale ($k_2=k_{COBE}
\simeq 10^{-3}h/\rm{Mpc}$) can be taken from $\Delta T/T$ due to the 
Sachs-Wolfe-effect (Sachs, Wolfe 1967):

\begin{equation}
\big\langle\left(\frac{\Delta T}{T}\right)^2\big\rangle_{10^0} =
\rm{S} + \rm{T} \simeq 1.1 \times 10^{-10},\;\;
\rm{S} = 0.04 \; \langle q^2\rangle_{10^0}.
\end{equation}
The relationship between the power spectrum at COBE scale and the variance 
of the $q$ potential averaged in $10^0$-angular-scale at the last 
scattering surface, involves an effective interval of the spectral 
wavelengths contributing to the latter:
\begin{equation}
\langle q^2\rangle_{10^0}^{1/2} = fq_{k_2}, \;\;\;\;
f^2\sim \ln \left( \frac{k_2}{k_{hor}} \right).
\end{equation}
To estimate $\rm{T/S}$, we will use the approximation formula (1) for $x_2 =
x_{COBE}$ (cf. eqs.(9), (10)):
\begin{eqnarray}
\frac{\rm T}{\rm S} \simeq -6n_T = 12\gamma &=& 
3\left( \frac{h_{k_2}}{q_{k_2}}\right)^2 \\
&=&\frac{6}{\sqrt{c^2 + x^2_2}}\left( 1-\frac{x_2}{\sqrt{c^2 + x^2_2}}\right).
\nonumber
\end{eqnarray}

Evidently, both normalizations determine essentially the corresponding 
$q_k$ amplitudes at the locations of cluster ($k_1$) and COBE ($k_2$) 
scales. Accordingly, the parameters $k_1$ and $f$ can slightly vary while
changing the local spectrum slopes at the respective wavelengths.  
However, when the deviation of the $q_k$ slopes from HZ is small (e.g. 
$c>5$, see eq.(8)) we can just identify the parameters $k_1$ and $f$ for 
CDM models with their sCDM values:
\[
k_1 \simeq 0.3h/\rm{Mpc}\;,\;\;\;\; f\simeq 1.26\;.
\]
Finally, comparing the two normalization conditions with the theoretical 
spectrum $q_k$ in eq.(5) we get the key 
equation for the relationship between $c$ and $k_{cr}$,
\begin{equation}
\left( \frac{q_{k_1}}{q_{k_2}} \right)^2 \simeq
D\left( 1 +\frac{\rm T}{\rm S}\right) ,
\end{equation}
which looks like an algebraic equation for finding $x_2$ by given $c$:
\begin{equation}
\left(1 + \frac{d(d+2x_2)}{c^2 + x^2_2}\right)^{3/2} \simeq D
\left(1 +\frac{6}{\sqrt{c^2+x^2_2}} - \frac{6x_2}{c^2+x^2_2}\right).
\end{equation}
Here $D \sim \sigma^2_8$, and 
$$
d = x_1 - x_2 \simeq 
\ln \left[ 300\left(\frac{c^2 + x_2^2}{c^2+ x_1^2}\right)^{1/12} 
\left(\frac{\sqrt{c^2+x_2^2} - x_2}{\sqrt{c^2 +x_1^2} - x_1} \right)^{2/3} 
\right].
$$
The $\lambda$-parameter is then obtained as
\[
\sqrt{\lambda} \simeq \frac{10^{-4}\sigma_8}{(c^2 + x_1^2)^{3/4}}.
\]

\begin{figure}
{\epsfxsize=9cm \epsfysize=8truecm
\epsfbox[25 10 226 226]{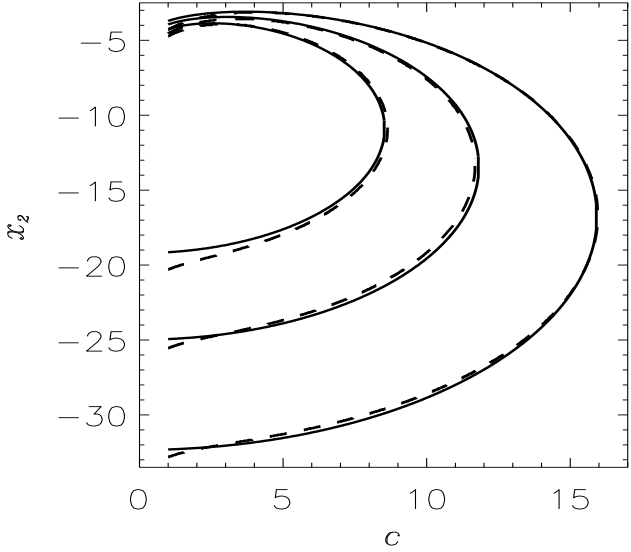}}
\vskip 0.2cm
\caption{The numerical solutions of eq.(17) (solid lines) and 
their approximations due the eq.(19) (dashed lines). The inner, central, 
and outer curves are for $D=0.2, 0.3, 0.4$, respectively.}
\end{figure}

Eq.(16) has a clear physical meaning:  the ratio of the S-spectral power 
at cluster and COBE scales is proportional to $\sigma_8^2$ and inversely
proportional to the fraction of the scalar mode contributing to the 
large-scale temperature anisotropy variance, S/(S+T). This simple 
argument makes eq.(16) independent of a particular way how it was 
obtained, just the proportionality coefficient should be taken properly.

Eq.(16) provides quite a general constraint on the fundamental
inflation spectra (both S and T) in a wide set of dark matter models
using only two basic measurement (the cluster abundance and large scale
$\Delta T/T$).  Actually, the DM information (a transfer function) is
contained in the $D$-coefficient which can be calculated using the same
equation (16) for a simple inflationary spectrum (e.q.  power-law) preserving
the given DM model.  E.g. for CDM with $h=0.5$ we have:
\begin{equation}
D\simeq \frac{0.6\sigma_8^2}{1-3.1\Omega_b}\; ,\;\;\;\;\;
\Omega_b <0.2.
\end{equation}

The solution of eq.(17) is shown in the plane $x_2 - c$ for D=0.2, 0.3, 0.4
(see Fig. 2). For the whole range $0.1<D<0.5$, it can be analytically 
approximated with a precision better than 10\% as follows:
\begin{equation}
\ln^2 \left(\frac{k_0}{k_{cr}}\right)\simeq E\left(c_0-c\right)\left(c
+c_1\right)\;,\;\;\;\; 2<c\leq c_0\;.
\end{equation}
Notice there exists not any solution of eq.(17) for high enough $c$ ($c>c_0$). 
We have found the following best fit coefficients $E,\;k_0[h/\rm{Mpc}]$ and 
$c_{0,1}$:
\[
E\simeq 1,\;\;\;\ln k_0\simeq 49D^2+1.3,
\]
\[
c_0\simeq 61D^2+6.2,\;\;\; c_1\simeq 44D^2+4.0.
\]
The tensor-mode-contribution is approximated similarly:
\begin{equation}
\frac{\rm T}{\rm S}\simeq \frac{2.53-4.3D}{(\ln k_{cr}+4.65)^{2/3}}
+\frac{1}{3}\;.
\end{equation}
Recall, $k_{cr}$ is measured in the units $h/\rm{Mpc}$.

\section{Discussion}

We have presented a new inflationary model predicting a near scale-invariant
spectrum of density perturbations and large amount of CGWs.  Our model is
consistent with COBE $\Delta T/T$ and cluster abundance data.  The
perturbation spectra depend on one free scale-parameter, $k_{cr}$, which can
be found in further analysis by fitting other observational data.  At the
location of $k_{cr}$, the CDP spectrum transients smoothly from the red
($k<k_{cr}$) to the blue ($k>k_{cr}$) parts (see eq.(5)).

By adjusting $k_{cr}$ with the galaxy cluster scale, we easily gain the
boosts in power (in comparison with sCDM) on both scales, large (voids and
superclusters) and small (quasars and $Ly_{\alpha}$ clouds).  Say, for
$k_{cr}=k_1,\; D=0.4$ and $\Omega_b =0.1$, we get $c \simeq 11$,
$\sqrt{\lambda}\simeq 1.6\cdot10^{-6}$ and
\begin{equation}
\rm{T/S} \simeq 0.7,\;\;\;\;
n_S\simeq 0.9
\end{equation}
at large scale ($\sim 1000 h^{-1}\rm{Mpc}$). The boost on Mpc-scale is about
$8\%$,
\[
\left( \frac{q_{10k_1}}{q_{k_1}}\right)^2
\simeq 1.08,
\]
which is really a lot when compared with the red spectrum (21)
extrapolated from large scales.
   
\begin{figure}
{\epsfxsize=9cm \epsfysize=10.5cm
\epsfbox[25 10 226 255]{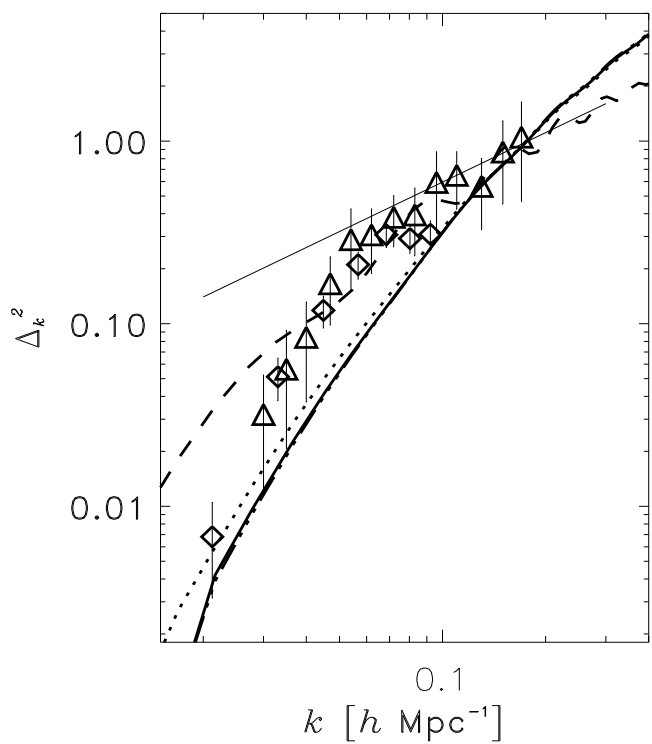}}
\vskip 0.2cm
\caption{Power spectrum of Abell-ACO clusters (triangles) according to 
Retzlaff et al.~(1998) and of APM clusters (diamonds) according to Tadros 
et al.(1998). The cluster power spectrum is reduced in amplitude 
to the galaxy level using a scale independent cluster bias factor $b_{cl}=2$. 
We compare with 4 theoretical models, sCDM (dash-tree-dotted line), LCDM 
with $\Omega_{\Lambda}=0.7$ (dashed), and the wing-like power 
spectra with parameters $c=5, 11$ as dotted and solid  
lines, respectiveley. The $\Lambda\lambda$ models with $c=5, 9, 11$ have 
mass variances $\sigma_8 = 0.4, 0.6, 0.7$, much 
lower than the unrealistic sCDM with $\sigma_8 = 1.1$. The $c=11$ curve 
lies almost on top of sCDM, the (not shown) $c=9$ curve lies between 
$c=5$ and $c=11$ curves. The solid straight line corresponds to the power 
law (22).}
\end{figure}

We conclude that the wing-like S-power-spectra similar to those in eq.(5),
can provide a simple solution to the cosmological problem in a 
matter-dominated universe (cp. also a similar spectrum of CDPs in Semig, 
M\"uller 1996).  Indeed, the spectrum power at the current dynamical scale is 
strongly suppressed by the present galaxy cluster abundance ($\sigma_8\simeq 0.6$),
whereas other `classical' observations persistently require a large CDP power
both on scales $\sim 100 h^{-1}\rm{Mpc}$ (the existence of large scale
structures) and $\le 1 h^{-1}\rm{Mpc}$ (QSOs, $Ly_{\alpha}$ forest, early
galaxy formation).

More of that, today we seriously discuss a nearly flat shape of the
dimensionless (linear) CDP-spectrum within the scale range encompassing
clusters and superclusters,
\begin{equation} 
\Delta_k^2\sim k^{(0.9 \pm 0.2)}\;,\;\;\;\;k \in (0.05, 0.2) h\; \rm{Mpc}^{-1},
\end{equation}
(with a break towards the HZ slope on higher scales) which stays in obvious
disagreement with the sCDM predictions.  The first arguments supporting 
eq.(22) came from the analysis of large-scale galaxy distribution (Guzzo et
al. 1991) and the discovery of large quasar groups (Komberg, Lukash 1994,
Komberg, Kravtsov, Lukash 1996).  Recent measurements of the galaxy cluster
power spectrum (Tadros, Efstathiou, Dalton 1998, Retzlaff et al. 1998)
brought a higher statistical support for eq.(22) (see Fig.3).

A possible explanation of eq.(22) from a theoretical point of view can be a
fundamental red power spectrum established on large scales; then the
transition to the spectrum (22) at $\sim 100h^{-1}\rm{Mpc}$ would be much easier
understood with help of a traditional modification of the transfer function
$T(k)$ (e.g.  for mixed hot+cold dark matter, see Mikheeva et al. 2000).  
A boost of power of the high-multipole CMB anisotropy (at $l\sim 220$) looks 
self-consistent with the
above argument on a slightly red S-spectrum at large scales.  The rediness
may be not too high, remaining in the range (0.9, 1). Notice any introduction 
of a blue power spectrum at large scale appears extremely unfavorable in this 
connection as it suppresses severely all the LSS effects at 
$\sim 100 h^{-1}\rm{Mpc}$ (provided the spectrum meets the $\sigma_8$ 
constraints).  

The main problem for the LSS in matter-dominated
models remains a low number of $\sigma_8$:  if $\sigma_8\le 0.6$, then the
first acoustic peak in $\Delta T$ cannot be as high as $\ge 70\;\mu K$ in any 
model with $\Omega_m=1$ (see Fig.4, cp. a similar conclusion for MDM models
in Mikheeva et al. 2000).  

\begin{figure} 
{\epsfxsize=9cm\epsfysize=10.5truecm 
\epsfbox[25 10 226 255]{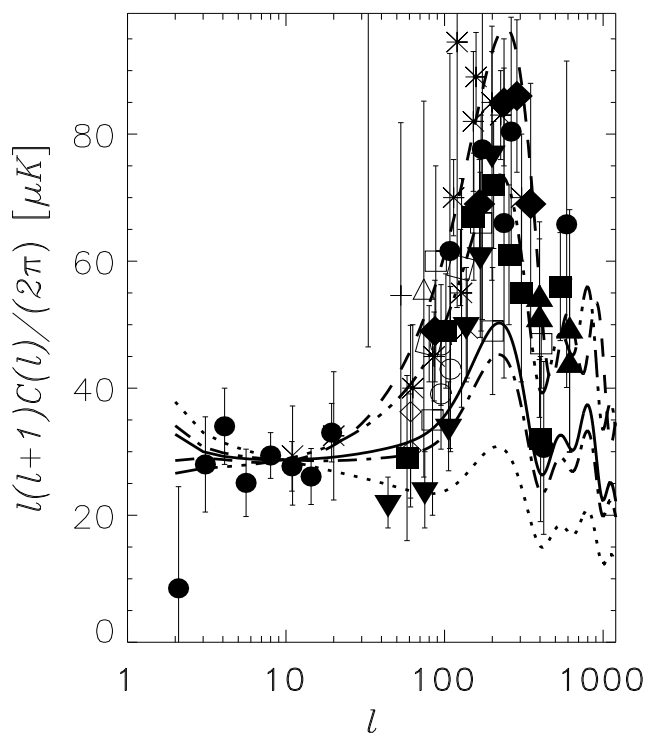}} 
\vspace{0.2cm}
\caption{Multipole moments of the 5 theoretical models, sCDM 
(dash-three-dotted), $\Lambda$CDM (dashed), and the wing-like power spectra 
with $c=5$ (dotted),  $c=9$ (dash-dotted), and  $c=11$ (solid line). The last 
curves have a tensor to scalar ratio T/S of 1.3, 0.9, and 0.7, respectively. 
For the calulation of the multipoles we used the Boltzmann code CMBFAST of 
Zaldarriaga, Seljak (1999). The observational data stem 
from the collection of experiments in Tegmark, Zaldarriaga (2000).}  
\end{figure}

While the red power spectrum could help in solving the LSS problem, it is
undesirable when continued to Mpc scale:  here we need an enhanced power of
CDPs to trigger the observed early structure formation.  This is the way to
see how a wing-like power spectrum can be of help. 

Summarizing, our arguments are very simple:

(i)  The wing-like S-power-spectra is a common feature of $\Lambda$-inflation;
 
(ii) An early galaxy formation period together with the
developed structure at $(10 - 100)h^{-1}\rm{Mpc}$ can be reconciled
(for models with standard cold or mixed dark matter) in the considered 
wing-like primordial spectrum of CDPs.  

\subsection*{Acknowledgments}
The work was partly supported by German Scientific Foundation (DFG-436 RUS
113/357/0) and INTAS grant (97-1192).

\bigskip
\section*{Appendix}
Here, we recall the basic properties of the parametric amplification 
effect and derive the cosmological perturbation spectra originating 
in $\Lambda\lambda$-inflation (Lukash, Mikheeva 2000).

The linear perturbations in the Friedmann geometry are irreducibly presented 
in terms of the uncoupled scalar and tensor parts:
$$
ds^{2} = (1+h_{00})\;dt^{2}+2ah_{0\alpha}\;dtdx^{\alpha}-a^{2}
(\delta_{\alpha\beta} + h_{\alpha\beta})\;dx^{\alpha}dx^{\beta},
\eqno(A.1)
$$
$$
\frac {1}{2}h_{\alpha\beta}=A\delta_{\alpha\beta}+B_{,\alpha\beta}+
G_{\alpha\beta},\;\;\;\; h_{0\alpha}=C_{,\alpha}, 
$$
where $G_{\alpha\beta}$ is a trace-free divergence-less tensor field  
($G^{\alpha}_{\alpha}=G^{\beta}_{\alpha,\beta}=0$), and potentials
$h_{00}$, $A$, $B$, $C$ are coupled to the perturbation of the scalar 
field $\varphi$ (which is also the source of the Friedmann geometry). The 
vector mode is not included here for the absence of proper source.

The gauge-invariant canonical 4-scalar determining the physical
scalar perturbations, $q=q(t,\vec{x})$, is uniquely fixed by the
appearance of the S-part of the perturbative Lagrangian similar to a massless 
field (Lukash 1980, 1996)\footnote{For simplicity, we assume the total field 
Lagrangian density in the form $\frac 12\varphi_{,i}\varphi^{,i}- V(\varphi)$ 
and the Hilbert minimal action for gravity.}:
$$
\delta L \equiv L(q,G_{\alpha\beta}) = \gamma q_{,i} q^{,i}+
\frac{1}{2}G_{\alpha\beta,\gamma}G^{\alpha\beta,\gamma}.
\eqno(A.2)    
$$
The relation of $q$ to the original potentials have the following form:
$$
\delta\varphi=\alpha\left(q+A\right),\;\;\;\;a^{2}\dot{B}+C=
 \frac{\Phi+A}{H}, 
$$
$$
\frac{1}{2}h_{00} = \gamma q + \left(\frac{A}{H}\right)^{.}, \;\;\;\;
\Phi=\frac{H}{a}\int a\gamma q dt,\eqno(A.3)
$$
$$
\frac{\delta\rho}{\rho+p} = \frac{\dot{q}}{H} - 3(q+A),\;\;\;\;
 4\pi G\delta\rho_{c} \equiv\gamma H\dot{q}=\frac{\triangle\Phi}{a^2},
$$
where $a$, $H$, $\gamma$, $\rho=\dot{{\varphi}^{2}}/2+V$,
$\alpha=\dot{\varphi}/H=\pm \sqrt{2\gamma}$ are solutions 
of the Friedmann equations of the background cosmological model, 
i.e. they are pure time dependent, 
and $\triangle\equiv\partial^{2}/\partial\vec{x}^{2}$ is the spatial 
Laplacian. Any two potentials taken from the triple $A$, $B$, $C$ 
are arbitrary functions of all coordinates specifying the gauge choice.

The equations of motions of the scalar and tensor fields propagating 
in the Friedmann Universe are two harmonic oscillators, 
$$
\ddot{q}+\left(3H+\frac{\dot{\gamma}}{\gamma}\right)\dot{q}-
\frac{\triangle}{a^2}q=0,\eqno(A.4)
$$
$$
\ddot{G_{\alpha\beta}}+3H\dot{G_{\alpha\beta}}-\frac{\triangle}{a^2}
G_{\alpha\beta}=0,\eqno(A.5)
$$
which reduce the problem of the generation of cosmological
perturbations to the well established parametric amplification
effect (particle creation in intensive gravitational fields).

The quantum-generated perturbations inflating outside the horizon
become frozen in time and can be treated as classical random
Gaussian fields with the variances
$$
\langle q^2 \rangle=\int^{k_{*}}_0 q^2_k \frac{dk}{k},\;\;\;\;
\langle G_{\alpha\beta}G^{\alpha\beta} \rangle=\int^{k_{*}}_0
h^2_k \frac{dk}{k},\eqno(A.6)
$$
where $k_{*}={\vert\eta\vert}^{-1}$ is the horizon-crossing
time ($\eta \equiv \int dt/a$). The power spectra are
given as usually in the limit $k\vert\eta\vert\ll 1$:
$$
q_{k}=\frac{k^{\frac{3}{2}}\vert\nu_{k}\vert}{2\pi a\sqrt{\gamma
}},\;\;\;\; 
h_{k}=\frac{k^{\frac{3}{2}}
\vert\nu_{k}^{\lambda}\vert}{\pi a},\eqno(A.7)
$$
by solving the respective Klein-Gordon equations for the
functions $\nu_{k}^{\lambda}(\eta)$ (see (A.4), (A.5)),
$$
\frac{d^{2}\nu_{k}^{(\lambda)}}{d\eta^{2}}+\left(k^{2}-
U^{(\lambda)}\right)\nu^{(\lambda)}_{k}=0,\eqno(A.8)
$$
with
$U=d^2\left(a\sqrt{\gamma}\right)/a\sqrt{\gamma}d\eta^2$,
$U^{\lambda} =d^2a/d\eta^2$, and the vacuum initial 
conditions,
$$
k\vert\eta\vert\gg 1: \;\;\;
\nu_{k}^{(\lambda)} = \frac{\exp(-ik\eta)}{\sqrt{2k}}. \eqno(A.9)
$$
In particular, for slow-roll inflation (e.g. the $\Lambda\lambda$-inflation) 
we have 
$$
\nu_{k}^{(\lambda)} \simeq
\frac{\exp(-ik\eta)}{\sqrt{2k}}\left(1-\frac{i}{k \eta}\right),
\eqno(A.10) 
$$
and
$$
q_k \simeq \frac{H}{2\pi\sqrt{2\gamma}},\;\;\;\; h_k \simeq
\frac{H}{\pi\sqrt{2}}, \;\;\;\; k=aH\simeq -1/\eta. \eqno(A.11)
$$
The $h_k$ spectrum takes into account both polarizations of CGWs.

In conclusion, we comment on eq.(13) recalling that $\Phi=0.6 q$ at the
matter-dominated stage (see (A.3)), and 
$$
\frac{\Delta T}{T}=\frac{\Phi}{3}=\frac{q}{5}
\eqno(A.12)
$$ 
at large angular scale ($>1^0$); both Newtonian and $q$ potentials 
in (A.12) are taken at the last-scattering surface.

\end{document}